\documentclass[twocolumn,nofootinbib,preprintnumbers,amsmath,amssymb]{revtex4}

\usepackage{graphicx}
\usepackage{dcolumn}
\usepackage{bm}
\usepackage{color}

\newcommand{\C}{\mathbb{C}}
\newcommand{\R}{\mathbb{R}}
\newcommand{\Z}{\mathbb{Z}}

\newcommand{\cG}{\mathcal{G}}
\newcommand{\cZ}{\mathcal{Z}}
\newcommand{\cH}{\mathcal{H}}
\newcommand{\cL}{\mathcal{L}}
\newcommand{\cU}{\mathcal{U}}
\newcommand{\bG}{\mathbf{G}}
\newcommand{\vC}{\vec{C}}
\newcommand{\vt}{\vec{\theta}}

\newcommand{\Pmic}{P_{\mathrm{mic}}}
\newcommand{\Pcan}{P_{\mathrm{can}}}

\newcommand{\Smic}{S_\textrm{mic}}
\newcommand{\Scan}{S_{\mathrm{can}}}

\newcommand{\bS}{\mathbf{\Sigma}}
\newcommand{\be}{\begin{equation}}
\newcommand{\ee}{\end{equation}}

\begin{document}


\title{Reconnecting statistical physics and combinatorics beyond ensemble equivalence}
\author{Tiziano Squartini}
\affiliation{IMT School for Advanced Studies, Lucca, Italy}
\author{Diego Garlaschelli}%
\affiliation{Lorentz Institute for Theoretical Physics, University of Leiden, The Netherlands}%

\date{\today}

\begin{abstract}
In statistical physics, the challenging combinatorial enumeration of the configurations of a system subject to hard constraints (microcanonical ensemble) is mapped to a mathematically easier calculation where the constraints are softened (canonical ensemble).
However, the mapping is exact only when the size of the system is infinite and if the property of ensemble equivalence (EE), i.e. the asymptotic identity of canonical and microcanonical large deviations, holds. 
For finite systems, or when EE breaks down, statistical physics is currently believed to provide no answer to the combinatorial problem.
In contrast with this expectation, here we establish exact relationships connecting conjugate ensembles in full generality, even for finite system size and when EE does not hold.
We also show that in the thermodynamic limit the ensembles are directly related through the matrix of canonical (co)variances of the constraints, plus a correction term that survives only if this matrix has an infinite number of finite eigenvalues.
These new relationships restore the possibility of enumerating microcanonical configurations via canonical probabilities, thus reconnecting statistical physics and combinatorics in realms where they were believed to be no longer in mutual correspondence.
\end{abstract}

\maketitle

For virtually any system consisting of a large number $n$ of interacting elements, an exact microscopic description is unfeasible and has to be replaced by a statistical one involving a probability distribution $P(\bG)$ over the microstates $\bG\in\cG$ of the system consistent with a certain number $K$ of macroscopic properties $\vC^*=(C^*_1,\dots,C^*_K)$ that are empirically accessible.
Here $\vC^*\equiv\vC(\bG^*)$ is the particular, known value of the $K$-dimensional vector $\vC(\bG)$ of macroscopic constraints, $\bG^*\in\cG$ is the (unknown) microstate, and $\cG$ is the set of allowed microstates.
For concreteness, throughout the paper we assume that $\bG$ only takes discrete values, so that $P(\bG)$ is a probability mass function. 
The principle of maximum entropy states that the optimal $P(\bG)$ is the one that maximizes the entropy functional
\be
 S[P]\equiv-\sum_{\bG\in\cG}P(\bG)\ln P(\bG),
\label{eq:entropy}
\ee
as such maximum-entropy probability is \emph{``maximally noncommittal with respect to missing information''}~\cite{Jaynes}.
The resulting theory is statistical physics, which describes large systems subject to a few macroscopic constraints such as total energy and total number of particles~\cite{Boltzmann1877,Gibbs1902}.

The usefulness of statistical physics crucially depends on whether one is able to calculate $P(\bG)$ explicitly.
Imposing \emph{hard} constraints, i.e. restricting oneself to the configurations $\bG$ such that $\vC({\bG})=\vC^*$, leads to the so-called \emph{microcanonical} maximum-entropy probability 
\be
\Pmic(\bG|\vC^*)=\Omega^{-1}_{\vC^*}\delta_{\vC^*,\vC(\bG)},
\label{eq:Pmic}
\ee
where $\Omega_{\vC^*}$ is the number of microstates compatible with the constraint $\vC^*$. 

The calculation of $\Omega_{\vC^*}$ requires complicated and often unknown combinatorial enumeration formulae, a mathematical challenge that makes the microcanonical ensemble generally intractable. 
A cornerstone of statistical physics, dating back to Gibbs~\cite{Gibbs1902} and known as `ensemble equivalence' (EE), is the assumption that, in the thermodynamic limit $n\to\infty$, the hard values $\vC^*$ of the constraints can be safely replaced by \emph{soft} (i.e. suitably fluctuating) values, since for large systems the fluctuations around the hard values are expected to vanish. 
The result is the mathematically tractable \emph{canonical} ensemble~\footnote{At this point, since we are considering a generic choice of constraints, in principle we need not make a distinction between the canonical ensemble (where the energy is the traditional soft constraint in statistical physics) and the grandcanonical ensemble (where the number $n$ of particles is an additional soft constraint). However, since later on we will write expressions that depend on $n$, for simplicity we restrict to the canonical ensemble.} 
where the constraints are enforced only as expected values, i.e.  $\langle\vC\rangle=\vC^*$. The maximization of the entropy yields in this case~\cite{Jaynes}
\be
\Pcan(\bG|\vt^*)=\frac{e^{-\vt^*\cdot\vC(\bG)}}{\sum_{\bG'\in\cG}e^{-\vt^*\cdot\vC(\bG')}}=\frac{e^{-\cH(\bG,\vt^*)}}{\cZ(\vt^*)},
\label{eq:Pcan}
\ee
where $\vt$ is a $K$-dimensional vector of Lagrange multipliers needed to control the expected value $\langle\vC|\vt\rangle\equiv\sum_{\bG\in\cG}\Pcan(\bG|\vt)\vC(\bG)$, $\cH(\bG,\vt)\equiv \vt\cdot\vC(\bG)=\sum_{k=1}^K\theta_k C_k(\bG)$ is (somewhat improperly) called the `Hamiltonian' function (the dot indicating the scalar product), $\cZ(\vt)=\sum_{\bG\in\cG}e^{-\vt\cdot\vC(\bG)}$ is the partition function,  and $\vt^*$ is the specific value such that 
\be
\langle\vC|\vt^*\rangle=\vC^*,
\label{eq:*}
\ee 
where we have defined the canonical ensemble average of a generic function $X(\bG)$ as
\be
\langle X|\vt\rangle\equiv\sum_{\bG\in\cG}\Pcan(\bG|\vt)X(\bG).
\label{eq:ensembleaverage}
\ee
Equivalently, the value $\vt^*$ is the one that maximizes the log-likelihood function $\cL^*(\vt)\equiv\ln \Pcan(\bG^*|\vt)$~\cite{mylikelihood,squartini2011} (see SI).

Note that eq.~\eqref{eq:*} fixes the relationship between the dual quantities $\vC^*$ and $\vt^*$.
We can therefore simplify the notation and set $\Pmic^*(\bG)\equiv\Pmic(\bG|\vC^*)$ and $\Pcan^*(\bG)\equiv\Pcan(\bG|\vt^*)$, the asterisk indicating that both probabilities are ultimately specified by the unique value $\vC^*$.
Similarly, after inserting eqs.~\eqref{eq:Pmic} and~\eqref{eq:Pcan} into eq.~\eqref{eq:entropy}, we can denote the resulting entropies of the two ensembles as
\begin{eqnarray}
\Smic^*&\!\!\equiv\!\!& S[\Pmic^*]=\ln\Omega_{\vC^*}
\label{eq:Smic}\\
\Scan^*&\!\!\equiv\!\!&S[\Pcan^*]=-\sum_{\bG\in\cG}\Pcan(\bG|\vt^*)\ln \Pcan(\bG|\vt^*).
\end{eqnarray}
An important equality, proven in SI and crucial for our later results, is the one between the canonical entropy and minus the \emph{maximized} log-likelihood $\cL^*(\vt^*)$:
\be
\Scan^*=-\ln \Pcan(\bG^*|\vt^*)=-\cL^*(\vt^*).
\label{eq:LS}
\ee
Throughout the paper, we assume that in the thermodynamic limit $\vt^*$ is not a critical point of a phase transition, i.e. it is in the interior of a thermodynamic phase.

Obviously, for finite $n$ the microcanonical and canonical ensembles are necessarily different. A mathematical formalization of this difference is discussed later. An informal example is the different physical interpretation of the two ensembles, the microcanonical one being viewed as a model for systems in energetic isolation (if the constraint is the total energy) and the canonical one as a model for systems in thermal equilibrium with a heat bath (if the Lagrange multiplier is the inverse temperature).
So it only makes sense to talk about EE in the thermodynamic limit, if relative energy fluctuations vanish. This already implies that the combinatorial problem cannot be solved via statistical physics when $n$ is finite.

In the thermodynamic limit, provided EE holds, the generally unfeasible calculations in the microcanonical ensemble can be replaced by easier calculations in the canonical ensemble.
For instance, most textbooks treat the microcanonical ensemble by relying on the approximation $\Omega_{\vC^*}\approx e^{\Scan^*}=1/\Pcan^*(\bG^*)$ where the symbol ``$\approx$'' means ``equal up to a factor assumed to be subexponential in $n$'' and the last equality follows from eq.~\eqref{eq:LS}.
To make a concrete example, for systems with given total energy  $E^*$ one often assumes $\Omega_{E^*}\approx e^{+\beta E^*}\cZ(\beta)$ where $\beta=(k_BT)^{-1}$ is the inverse temperature, $k_B$ is Boltzmann's constant and $T$ is the temperature. 
Stated more rigorously, this assumption is
\be
\Omega_{\vC^*}= e^{\Scan^*-o(n)}=\frac{e^{-o(n)}}{\Pcan^*(\bG^*)}
\label{eq:approx}
\ee
(where $o(n)$ denotes a quantity that, if divided by $n$, vanishes as $n$ diverges) or equivalently 
\be
\lim_{n\to\infty}\frac{\ln\Omega_{\vC^*}}{n}=-\lim_{n\to\infty}\frac{\ln\Pcan^*(\bG^*)}{n}=\lim_{n\to\infty}\frac{\Scan^*}{n}.
\label{eq:map}
\ee
Therefore, in presence of EE, the enumeration problem required to calculate the `hard' microcanonical quantity $\Omega_{\vC^*}$ can be mapped exactly, in the thermodynamic limit, to a `softened' canonical problem requiring only the calculation of $\Pcan^*(\bG^*)$ or $\Scan^*$. This establishes a tight connection between statistical physics and combinatorics.
An equivalent condition is
\be
\Pmic^*(\bG^*)=\Omega^{-1}_{\vC^*}= e^{o(n)-\Scan^*}=\Pcan^*(\bG^*)e^{o(n)},
\label{eq:Papprox}
\ee
or ultimately
\be
\lim_{n\to\infty}\frac{1}{n}\ln\frac{\Pmic^*(\bG^*)}{\Pcan^*(\bG^*)}=0,
\label{eq:ultimately}
\ee
which essentially states that EE corresponds to the identity of the large deviation properties of the two conjugate ensembles at rate $n$, i.e. the two probability measures are \emph{exponentially equivalent} at rate $n$. 

When the size $n$ of the system is finite, or when the assumption of EE as stated by eq.~\eqref{eq:ultimately} is violated, the two ensembles are different and the enumeration problem cannot be directly solved via a canonical calculation.
Indeed, while most statistical physics textbooks still convey the idea that EE is expected to hold in very general circumstances, several examples of the breakdown of EE in a diverse range of systems have actually been provided over the last decades~\cite{Ellis2000,Blume1971,Barre2001,Ellis2004,LyndenBell1999,Chavanis2003,DAgostino2000,Barre2007}.
A recently investigated class of such systems is random networks with topological constraints~\cite{RS,RY,Chatterjee,Roccaverde,ginestra,myunbiased,mybreaking,mymodular}, to which we will briefly refer later in this paper.
For systems with nonequivalent ensembles, it has been shown that microcanonical and canonical calculations of various properties no longer agree. 
Therefore, for these systems, the connection between statistical physics and combinatorics appears essentially lost, just like for  systems with finite size.

At this point, an open problem of practical relevance emerges. Even accepting that the useful mapping to the canonical ensemble is only justified in the regime of EE, how to rigorously determine whether EE holds, without having to calculate the microcanonical ensemble? In fact, if EE can only be verified \emph{after} having calculated \emph{both} canonical and microcanonical properties and compared them, then the practical usefulness of the canonical ensemble evaporates, because one will have to trust the latter only when the microcanonical one is also tractable.

A second open problem, related to the previous one but of more fundamental theoretical importance and generality, should also be highlighted. Is it possible, in absence of EE and/or when the size of the system is finite, to restore some (possibly modified) rigorous mathematical correspondence between the canonical and microcanonical ensembles, so that the intractability of the latter can still be circumvented by performing feasible calculations in the former?
In other words, is there a general and stronger mathematical relationship connecting conjugate statistical ensembles, irrespective of whether EE holds and possibly for finite system size?

We now provide definite solutions to these two problems.
One of our main results, proven in SI, is the following exact formula, which is valid in general and reduces the calculation of $\Omega_{\vC^*}$ to purely canonical quantities:
\be
\Omega_{\vC^*}\!=\!\Pmic^{-1}(\bG^* | \vC^*)=\!\!\int_{-\vec{\pi}}^{+\vec{\pi}}\!\! \frac{d\vec{\psi}}{(2\pi)^K}\,\Pcan^{-1}(\bG^* | \vt^*\!+\!i\vec{\psi})
\label{eq:connection}
\ee
where $\vec{\pi}\equiv(\pi,\dots,\pi)$ is the $K$-dimensional vector with all components equal to $\pi$ and the argument $\vt$ of $\Pcan(\bG^* | \vt)$ has been extended to complex values.
The above equation establishes a seemingly esoteric, but exact relationship connecting the canonical and microcanonical probabilities. When the integral in eq.~\eqref{eq:connection} can be carried out explicitly, the microcanonical ensemble can be calculated \emph{exactly} from an extension of the canonical ensemble to the complex domain.
Remarkably, the formula is valid \emph{even for finite $n$ and even when EE breaks down}, thus extending the usefulness of the canonical ensemble significantly and unveiling a tighter and more general connection between `hard' combinatorial enumeration and `soft' statistical physics.

Due to the importance of the above result, we may wonder whether eq.~\eqref{eq:connection}, which in principle needs to be evaluated case by case and is therefore not very transparent, takes some more explicit and general form, independently of the model specification.
As we show in SI, this is indeed possible.
Before showing the result, we need some precautions. Note that, for certain choices of $\vt^*$ or $\vC^*$, some of the $K$ components of $\vC^*$ may be dependent on the other ones (see SI for details). In such a case, there is only an effective number $\tilde{K}\le K$ of independent constraints. 
Upon ordering the entries of $\vC$ in such a way that the first $\tilde{K}$ ones are independent, we define the reduced $\tilde{K}\times\tilde{K}$ matrix $\tilde{\bS}^*$ of the canonical (co)variances, evaluated at $\vt^*$, of the $\tilde{K}$ independent constraints.
The entries of $\tilde{\bS}^*$ are 
\be
\tilde{\Sigma}^*_{k,l}=\textrm{Cov}[C_k,C_l]_{\vt^*}\qquad k,l=1,\dots,\tilde{K}
\label{eq:sigma}
\ee
and its eigenvalues, which are all strictly positive, are denoted as $\{\lambda^*_k\}_{k=1}^{\tilde{K}}$.
The strict positivity of these eigenvalues is a consequence of the semi-positive definiteness of covariance matrices, together with the fact that all the $\tilde{K}$ rows of $\tilde{\bS}^*$ are independent, as follows from the independence of the corresponding $\tilde{K}$ constraints.
Armed with the above notation, we can carry out a saddle-point~\cite{saddlepoint} calculation (see SI) to obtain the following asymptotic expansion for $\Omega_{\vC^*}$:
\be
\Omega_{\vC^*}=
\frac{e^{\Scan^*}T^*}{\sqrt{\det(2\pi\tilde{\bS}^*)}}
\label{eq:omegaapprox}
\ee
where  
\be
T^*\equiv\prod_{k=1}^{\tilde{K}}\left[1+O\left(1/\lambda^*_k\right)\right].
\label{eq:T}
\ee
Equation~\eqref{eq:omegaapprox} makes the leading term of eq.\eqref{eq:connection} manifest and renders the connection between the hard enumeration problem and its softened dual more explicit. In particular, upon comparison with eq.\eqref{eq:approx}, it shows that an important role is played not only by the canonical entropy $\Scan^*$, but also by the canonical covariances $\tilde{\bS}^*$.
We will discuss the asymptotic behaviour of $T^*$ later.
For the moment, we note that if no $n$-dependence is assumed in the problem, i.e. question is counting how many configurations are compatible with a fixed constraint $\vC^*$ for a given finite value of $n$, then eqs.~\eqref{eq:omegaapprox} and~\eqref{eq:T} already provide the corresponding answer. 
However, the usual problem in combinatorics is that of assuming a certain $n$-dependence of the configuration space and of the enforced constraint. In this case, the counting question needs an asymptotic answer for large $n$. This is precisely what we are going to investigate at the end, when moving to the thermodynamic limit. 

We now show how the above results relate to the property of ensemble (non)equivalence and how they can be exploited to generalize eqs.\eqref{eq:approx},~\eqref{eq:map},~\eqref{eq:Papprox} and~\eqref{eq:ultimately} to the case of finite system sizes and/or nonequivalent ensembles.
We start from the definition of the relative entropy (or Kullback-Leibler divergence) between the canonical and microcanonical probabilities~\cite{touchette2014,mybreaking}:
\be
\Delta^*\equiv S[\Pmic^*||\Pcan^*]=\sum_\bG\Pmic^*(\bG)\ln\frac{\Pmic^*(\bG)}{\Pcan^*(\bG)}.
\label{eq:KL}
\ee
The relative entropy rigorously quantifies the additional uncertainty contained in $\Pcan^*(\bG)$ as compared to $\Pmic^*(\bG)$. This additional uncertainty is due to the fluctuating nature of the constraints in the canonical ensemble.
An important result, proven in ref.~\cite{mybreaking}, is that $\Delta^*$ is only determined by the \emph{local} values $\Pcan^*(\bG^*)$ and $\Pmic^*(\bG^*)$ achieved by any of the configurations $\bG^*$ that realize the sharp value $\vC^*$ of the constraints:
\be
\Delta^*=\ln\frac{\Pmic^*(\bG^*)}{\Pcan^*(\bG^*)}=\ln\frac{1}{\Omega_{\vC^*}\Pcan^*(\bG^*)}.
\label{eq:Qcan}
\ee

For finite $n$, $\Pmic^*$ and $\Pcan^*$ are different and $\Delta^*>0$, in line with our previous discussion of the distinction between thermal equilibrium and energetic isolation. 
However, we can now quantify $\Delta^*$ exactly by inserting eq.\eqref{eq:connection} into eq.\eqref{eq:Qcan}:
\begin{eqnarray}
\Delta^*&=&-\ln\int_{-\vec{\pi}}^{+\vec{\pi}}\!\! \frac{d\vec{\psi}}{(2\pi)^K}\,\frac{\Pcan(\bG^* | \vt^*)}{\Pcan(\bG^* | \vt^*\!+i\vec{\psi})}\nonumber\\
&=&-\ln\int_{-\vec{\pi}}^{+\vec{\pi}}\!\! \frac{d\vec{\psi}}{(2\pi)^K}
\langle e^{i\vec{\psi}\cdot(\vC^*-\vC)}|\vt^*\rangle,
\label{eq:Phigeneral}
\end{eqnarray}
where we have used eq.~\eqref{eq:ensembleaverage} with $X(\bG)=e^{i\vec{\psi}\cdot[\vC^*-\vC(\bG)]}$ (see SI for details).
Equation~\eqref{eq:Phigeneral} is valid in general (again, even for finite $n$ and when EE breaks down) and provides a way to calculate the relative entropy exactly \emph{as a function of only the canonical probability} (extended to the complex domain).

We now consider the thermodynamic limit.
Inserting the asymptotic result~\eqref{eq:omegaapprox} into eq.~\eqref{eq:Qcan} leads to
\begin{eqnarray}
\Delta^*&=&\ln\frac{\sqrt{\det(2\pi\tilde{\bS}^*)}}{T^*},
\label{eq:deltaT}
\end{eqnarray}
where we have used eq.~\eqref{eq:LS}.
It is now useful to re-express the above result in terms of a suitable relative entropy \emph{density}.
Indeed, the definition of EE in the measure sense~\cite{touchette2014} is the vanishing of the relative entropy \mbox{$n$-density} as $n\to\infty$, i.e. 
\be
\lim_{n\to\infty}\frac{\Delta^*}{n}=0.
\label{eq:measure}
\ee
Using eq.~\eqref{eq:Qcan}, we indeed see that the above condition is the same as eq.~\eqref{eq:ultimately}. When EE breaks down, i.e. when the above limit is strictly positive, then eqs.~\eqref{eq:approx},~\eqref{eq:map},~\eqref{eq:Papprox} and~\eqref{eq:ultimately} no longer hold, and the nexus between combinatorics and statistical physics is apparently severed. 
However, this turns out not to be the case, because a nontrivially modified version of all those equations is still in place, as we now show.

Let us slightly change our perspective and look for an increasing positive sequence $\alpha_n$ such that, irrespective of whether EE holds, the following limit is finite:
\be
\delta^*_{\alpha_\infty}\equiv\lim_{n\to\infty}\frac{\Delta^*}{\alpha_n},\qquad\delta^*_{\alpha_\infty}\in(0,+\infty).
\label{eq:delta}
\ee
We call $\delta^*_{\alpha_\infty}$ the \emph{limiting ${\alpha_n}$-density} of $\Delta^*$.
We can then restate the definition of EE in the measure sense by saying that the ensembles are equivalent iff $\alpha_n=o(n)$.
We will return on this point later.
An important result, following from eq.\eqref{eq:deltaT} and proven in SI, is
\be
\delta^*_{\alpha_\infty}=\lim_{n\to\infty}\frac{\ln\sqrt{\det(2\pi\tilde{\bS}^*)}}{\alpha_n}+\tau^*_{\alpha_\infty}.
\label{eq:deltalimit}
\ee
where
\be
\tau^*_{\alpha_\infty}=-\lim_{n\to\infty}\frac{\ln T^*}{\alpha_n}
\label{eq:tau}
\ee
is zero, unless there is an infinite (growing at least like $\alpha_n$) number of eigenvalues of $\tilde{\bS}^*$ that have a finite limit as $n\to\infty$ (see SI). This key result finally allows us to discuss the asymptotic behaviour of $T^*$ in terms of its effects on $\tau^*_{\alpha_\infty}$.

Whenever the number of constraints is finite, then $\tau^*_{\alpha_\infty}=0$ irrespective of the value of the eigenvalues of $\tilde{\bS}^*$. This applies to most traditional situations in statistical physics, where there are only a handful of constraints including properties like total energy, total number of particles (possibly of different chemical species), etc. In the context of network ensembles, this also applies to e.g. random graphs with a finite number of global constraints, like total number of links and/or triangles and/or wedges~\cite{RS,RY,Chatterjee,Roccaverde}.

Also, whenever the canonical fluctuations of all constraints grow as the size of the system grows, then $\tau^*_{\alpha_\infty}=0$ irrespective of the number of constraints. Note that this is also a very natural situation, since the constraints are typically chosen to be extensive quantities whose mean and variance grows with $n$. Besides the traditional examples mentioned above, this includes systems with \emph{local} (i.e. particle-specific) constraints, such as dense random graphs with given degrees, for which $\alpha_n=n\ln n$, $K=\tilde{K}=n$, and $\tilde{\bS}^*$ has diagonal entries equal to the (diverging) variances of the (diverging) degrees. A rigorous mathematical proof that this class of graphs obeys eq.~\eqref{eq:deltalimit} with $\tau^*_{\alpha_\infty}=0$ is left for a separate paper~\cite{mycovariance}.

If the number of constraints is infinite and, roughly speaking, the canonical fluctuations of all but a finite number of constraints grow as the size of the system grows, then all but a finite number of eigenvalues will diverge, and again $\tau^*_{\alpha_\infty}=0$. This is a useful criterion in hybrid situations where $\tilde{\bS}^*$ has both finite and infinite eigenvalues.
So, in order to have a finite correction $\tau^*_{\alpha_\infty}>0$ surviving in eq.~\eqref{eq:deltalimit}, we need an infinite number of finite constraints. We expect this to occur in sparse random graphs with given degrees, for which $\alpha_n=n$, $K=\tilde{K}=n$, and $\tilde{\bS}^*$ is nearly diagonal with diagonal entries equal to the (finite) variances of the (finite) degrees~\cite{mycovariance}.
When $\tau^*_{\alpha_\infty}>0$, the value of $\delta^*_{\alpha_\infty}$ has to be calculated independently from eq.~\eqref{eq:deltalimit} for the specific model, using either the explicit form of $\Omega_{\vC^*}$ (if known) or carrying out the integral in  eq.~\eqref{eq:Phigeneral} explicitly and then taking the limit~\eqref{eq:delta}.

Note that, in our approach, whether $\tau^*_{\alpha_\infty}$ is zero has nothing to do with EE or its breakdown. By definition of $\alpha_n$, $\delta^*_{\alpha_\infty}$ is always strictly positive, irrespective of the value of $\tau^*_{\alpha_\infty}$. Whether EE holds depends on a different property, namely whether $\alpha_n=o(n)$. Our results above enable the calculation of $\delta^*_{\alpha_\infty}$, and hence $\alpha_n$, in terms of purely canonical quantities. This allows checking whether the ensembles are equivalent, without having to calculate microcanonical quantities.

We can now exploit our results in order to provide exact and more general replacements for eqs.~\eqref{eq:approx},~\eqref{eq:map},~\eqref{eq:Papprox} and~\eqref{eq:ultimately}, valid even when EE is violated.
Note that eq.~\eqref{eq:delta} implies
\be
\Delta^*={\alpha}_n \delta^*_{\alpha_\infty}+o({\alpha}_n),
\label{eq:deltadelta}
\ee
where $\delta^*_{\alpha_\infty}$ is given by eq.~\eqref{eq:deltalimit}. 
This implies that eq.~\eqref{eq:Qcan} can be rearranged as follows:
\begin{eqnarray}
\Omega_{\vC^*}&=&e^{\Scan^*-\Delta^*}\nonumber\\
&=&e^{\Scan^*-\alpha_n \delta^*_{\alpha_\infty}-o(\alpha_n)}\nonumber\\
&=&\frac{e^{-\alpha_n \delta^*_{\alpha_\infty}-o(\alpha_n)}}{\Pcan^*(\bG^*)}.
\label{eq:newapprox}
\end{eqnarray}
The above expression is the correct extension of eq.~\eqref{eq:approx}.
We see the appearance of the extra term $\delta^*_{\alpha_\infty}$ of finite order.
As a consequence, eq.~\eqref{eq:map} should be generalized to
\begin{eqnarray}
\lim_{n\to\infty}\frac{\ln\Omega_{\vC^*}}{\alpha_n}&=&
\lim_{n\to\infty}\frac{\Scan^*}{\alpha_n}-\delta^*_{\alpha_\infty},\label{eq:newmap}\\
&=&
\lim_{n\to\infty}\frac{\Scan^*-\ln\sqrt{\det(2\pi\tilde{\bS}^*)}}{\alpha_n}-\tau^*_{\alpha_\infty}\nonumber
\end{eqnarray}
which represents a novel, corrected identify mapping the enumeration problem to the calculation of the canonical entropy \emph{and} the relative entropy or the covariance matrix of the constraints. 
Correspondingly, eqs.~\eqref{eq:Papprox} and~\eqref{eq:ultimately} should be replaced by 
\be
\Pmic^*(\bG^*)= 
\Pcan^*(\bG^*)\,e^{\Delta^*}=\Pcan^*(\bG^*)\,e^{\alpha_n \delta^*_{\alpha_\infty}+o(\alpha_n)}
\label{eq:Papprox2}
\ee
and
\be
\lim_{n\to\infty}\frac{1}{\alpha_n}\ln\frac{\Pmic^*(\bG^*)}{\Pcan^*(\bG^*)}=\delta^*_{\alpha_\infty}
\label{eq:newultimately}
\ee
respectively. 

Equations~\eqref{eq:newapprox},~\eqref{eq:newmap},~\eqref{eq:Papprox2} and~\eqref{eq:newultimately} restore an exact, nontrivially modified connection between statistical ensembles, valid irrespective of whether they are equivalent. 
In presence of EE, i.e. if $\alpha_n=o(n)$, then $\alpha_n \delta^*_{\alpha_\infty}+o(\alpha_n)=o(n)$ and these expressions reduce to eqs.~\eqref{eq:approx},~\eqref{eq:map},~\eqref{eq:Papprox} and~\eqref{eq:ultimately} respectively. 
By contrast, if EE does not hold, then the factor neglected in eq.~\eqref{eq:approx} is actually (at least) exponential in $n$, i.e. the volume of microcanonical configurations is exponentially smaller than estimated in that equation.
Remarkably, our results restore, in a properly modified way, the possibility of conveniently expressing microcanonical quantities in terms of purely canonical ones, even beyond the regime of EE. As a byproduct, this shows that it is indeed possible to check for EE (i.e. whether $\alpha_n=o(n)$) without having to calculate the microcanonical ensemble.

We finally note that all choices of $\alpha_n$ with the same leading order are equivalent. This can be exploited to introduce a sort of `natural' $\tilde{\alpha}_n$ defined as
\be
\tilde{\alpha}_n=\ln\sqrt{\det(2\pi\tilde{\bS}^*)}
\label{eq:alphatilde}
\ee
(note that the term on the right is $n$-dependent, despite the absence of an explicit symbol ``$n$'' in our notation).
The above choice is admissible, because it always ensures $\delta^*_{\tilde{\alpha}_\infty}\in(0,+\infty)$. 
Moreover, it allows to rewrite a number of expressions in more compact form.
In particular, checking for EE simply reduces to checking whether $\tilde{\alpha}_n=o(n)$, i.e.
\be
\ln\sqrt{\det(2\pi\tilde{\bS}^*)}=o(n)\iff \mathrm{EE}.
\ee
This explicit and simple criterion is an important byproduct of the results presented in this paper.
Moreover, if there is at most a finite number of eigenvalues of $\tilde{\bS}^*$ that have a finite limit as $n\to\infty$, then $\tau^*_{\tilde{\alpha}_\infty}=0$ and we can rewrite eqs.~\eqref{eq:deltalimit} and~\eqref{eq:deltadelta} as
\begin{eqnarray}
\delta^*_{\tilde{\alpha}_\infty}&=&1,\label{eq:deltatilde}\\
\Delta^*&=&{\tilde{\alpha}}_n +o({\tilde{\alpha}}_n).
\label{eq:newdeltadelta}
\end{eqnarray}
Similarly, eqs.~\eqref{eq:newapprox} and \eqref{eq:newmap} become
\begin{eqnarray}
\Omega_{\vC^*}
=e^{\Scan^*-\tilde{\alpha}_n-o(\tilde{\alpha}_n)},
\label{eq:newnewapprox}\\
\lim_{n\to\infty}\frac{\ln\Omega_{\vC^*}}{\tilde{\alpha}_n}=
\lim_{n\to\infty}\frac{\Scan^*}{\tilde{\alpha}_n}-1,\label{eq:newnewmap}
\end{eqnarray}
while eqs.~\eqref{eq:Papprox2} and~\eqref{eq:newultimately} become
\begin{eqnarray}
\Pmic^*(\bG^*)=
\Pcan^*(\bG^*)\,e^{\tilde{\alpha}_n +o(\tilde{\alpha}_n)},\\
\lim_{n\to\infty}\frac{1}{\tilde{\alpha}_n}\ln\frac{\Pmic^*(\bG^*)}{\Pcan^*(\bG^*)}=1
\label{eq:newnewultimately}
\end{eqnarray}
respectively. Equations~\eqref{eq:deltatilde}-\eqref{eq:newnewultimately} are an elegant and compact summary of our results in the case $\tau^*_{\tilde{\alpha}_\infty}=0$.

To summarize, in this paper we have established new general connections between conjugate ensembles.
So far, the microcanonical and canonical ensembles have been regarded as dual representations of the same system that are valid under different physical conditions (such as energetic isolation and thermal equilibrium) and only coincide under certain assumption (thermodynamic limit and ensemble equivalence). 
By contrast, we have found that the relationship between the two ensembles is much stronger than mere conjugacy or duality, as it can be formulated in terms of mathematical identities that allow the two ensembles can be calculated from each other in full generality, even when neither the thermodynamic limit nor the assumption of ensemble equivalence are in order.
Our results significantly expand the usefulness of the canonical ensemble as a tool to carry out \emph{exact} physical and combinatorial calculations that are unfeasible in the microcanonical one even when EE breaks down and/or for finite system size.
In particular, we found new and very general enumeration formulae (both exact and asymptotic) that can be applied to a variety of (possibly unsolved) combinatorial problems. 
What is exciting from a physical point of view is that the enumeration of all the configurations with a given hard value of the constraints can be re-expressed in terms of the conjugate canonical distribution with softened constraints, extended to the complex domain. 
In the thermodynamic limit, the two ensembles are connected via the relative entropy density, which we found to be proportional to the logarithm of the determinant of the matrix of canonical covariances of the constraints, plus a possible correction term that is nonzero only if that matrix has an infinite number of finite eingenvalues.
These results offer new insight into the foundations of statistical physics and its connections to combinatorial enumeration. 
Moreover, they significantly expand the toolkit for studying systems with nonequivalent ensembles.

\acknowledgements
The authors acknowledge fruitful discussions with Frank den Hollander and  Andrea Roccaverde.

\clearpage
\newpage
\appendix
\setcounter{equation}{0}
\renewcommand{\theequation}{S\arabic{equation}}
\section*{SUPPLEMENTARY INFORMATION}
{\center
accompanying the paper\\
\emph{``Reconnecting statistical physics and combinatorics beyond ensemble equivalence''}\\
by T. Squartini and D. Garlaschelli\\
}

\subsection*{The likelihood function}
We first recall the result, shown e.g. in~\cite{mylikelihood,squartini2011}, that the value $\vt^*$ that realizes $\langle\vC|\vt^*\rangle=\vC^*$ in eq.~\eqref{eq:*} can be equivalently defined as the value that maximizes the \emph{log-likelihood function}
\begin{eqnarray}
\cL^*(\vt)&\equiv&\ln\Pcan(\bG^*|\vt)\nonumber\\
&=&-\vt\cdot\vC^*-\ln \cZ(\vt)
\label{eq:likelihood}
\end{eqnarray}
(where the asterisk now indicates dependence of $\cL$ on $\bG^*$ and hence on $\vC^*$, but \emph{not} on $\vt^*$).
In other words,
\be
\vt^*=\textrm{argmax}_{\vt}\left\{\cL^*(\vt)\right\}.
\label{eq:argmax}
\ee

To prove the above result, we calculate the first partial derivatives of $\cL^*$
\begin{eqnarray}
\frac{\partial}{\partial\theta_k}\cL^*(\vt)&=&-C^*_k-\frac{1}{\cZ(\vt)}\frac{\partial }{\partial\theta_k}\cZ(\vt)\nonumber\\
&=&-C^*_k+\sum_{\bG\in\cG}C_k(\bG)\frac{e^{-\vt\cdot\vC(\bG)}}{\cZ(\vt)}\nonumber\\
&=&\langle C_k|\vt\rangle-C^*_k
\label{eq:gradient}
\end{eqnarray}
and note that eq.~\eqref{eq:*} implies that $\vt^*$ is such that
\be
\frac{\partial}{\partial\theta_k}\cL^*(\vt)\bigg|_{\vt=\vt^*}=\langle C_k|\vt^*\rangle-C^*_k=0\quad\forall k.
\label{eq:nogradient}
\ee
In other words, $\vt^*$ is a stationary point for $\cL^*$. It is also the \emph{only} such point. 

To prove that $\vt^*$ is also a global maximum, we consider the second derivatives
\begin{eqnarray}
\frac{\partial^2}{\partial\theta_k\partial\theta_l}\cL^*(\vt)\!&=&\!\!\sum_{\bG\in\cG}\!C_k(\bG)\frac{e^{-\vt\cdot\vC(\bG)}}{\cZ(\vt)}\sum_{\bG\in\cG}\!C_l(\bG)\frac{e^{-\vt\cdot\vC(\bG)}}{\cZ(\vt)}\nonumber\\
&&-\sum_{\bG\in\cG}\!C_k(\bG)C_l(\bG)\frac{e^{-\vt\cdot\vC(\bG)}}{\cZ(\vt)}\nonumber\\
&=&\langle C_k|\vt\rangle\langle C_l|\vt\rangle-\langle C_kC_l|\vt\rangle\nonumber\\
&=&-\textrm{Cov}[C_k,C_l]_{\vt}
\end{eqnarray}
and evaluate them at the point $\vt=\vt^*$, to get
\be
\frac{\partial^2}{\partial\theta_k\partial\theta_l}\cL^*(\vt)\bigg|_{\vt=\vt^*}=-\textrm{Cov}[C_k,C_l]_{\vt^*}=-\Sigma^*_{k,l}
\label{eq:sigma2}
\ee
where $\Sigma^*_{k,l}$, as in eq.~\eqref{eq:sigma}, denotes the covariance of the constraints $C_k$ and $C_l$ under the canonical probability $\Pcan^*$.
The above result implies that, at the particular point $\vt^*$, the Hessian matrix $\mathbf{\Lambda}^*$ of second derivatives of $\cL^*$ is equal to minus the covariance matrix $\bS^*$ of the constraints.

Since covariance matrices are positive semi-definite (i.e. their eigenvalues are all non-negative), $\mathbf{\Lambda}^*$ is negative semi-definite (i.e. its eigenvalues are all non-positive). 
The only case when some eigenvalues of $\bS^*$ are zero is when some of the $K$ constraints are not linearly independent of the other constraints. 
This means that there are some redundant constraints, i.e. some of the imposed properties are linear combinations of other imposed properties. 
While this situation is generally not encountered, it may happen in certain phases of some models. 
If we exclude this circumstance, $\bS^*$ is positive definite and $\mathbf{\Lambda}^*$ is negative definite, which implies that $\vt^*$ is a maximum of $\cL^*(\vt)$. Since it is the only stationary point, it is necessarily a global maximum for $\cL^*(\vt)$.

By contrast, in the case when only an effective number $\tilde{K}<K$ if constraints are linearly independent of each other, there will be degenerate maxima for $\cL^*(\vt)$. If we order the constraints in such a way that the first $\tilde{K}$ ones are all linearly independent, while the following $K-\tilde{K}$ ones are dependent on the first $\tilde{K}$ ones, then $\vt^*$ will be a global maximum of $\cL^*(\vt)$ in the subspace spanned by the first $\tilde{K}$ Lagrange multipliers, while along the other $K-\tilde{K}$ directions $\cL^*(\vt)$ will be constant. Correspondingly, there will be a $(K-\tilde{K})$-dimensional family of degenerate solutions $\vt^*$ to eq.~\eqref{eq:*}.

In either case, the above discussion proves eq.~\eqref{eq:argmax}, which is a crucial result allowing us to prove a variety of other identities below.

\subsection*{Relation between canonical entropy and likelihood}
We here prove the important result that the maximized log-likelihood $\cL^*(\vt^*)$ coincides with $-\Scan^*$. This is easily demonstrated as follows:
\begin{eqnarray}
\cL^*(\vt^*)&=&\ln\Pcan^*(\bG^*)\nonumber\\
&=&-\vt^*\cdot\vC(\bG^*)-\ln \cZ(\vt^*)\nonumber\\
&=&-\vt^*\cdot\vC^*-\ln \cZ(\vt^*)\nonumber\\
&=&-\vt^*\cdot\langle\vC|\vt^*\rangle-\ln \cZ(\vt^*)\nonumber\\
&=&\langle \ln\Pcan^*(\bG)|\vt^*\rangle\nonumber\\
&=&-\sum_\bG \Pcan^*(\bG)\ln\Pcan^*(\bG)\nonumber\\
&=&-\Scan^*.\label{eq:L=-S}
\end{eqnarray}
The above result proves eq.~\eqref{eq:LS} and will be useful multiple times in the following.

\subsection*{Exact relation between $\Omega_{\vC^*}$ and $\Pcan^*$}
Here we prove one of our main results, i.e. the exact formula~\eqref{eq:connection} connecting $\Omega_{\vC^*}$, $\Pmic^*$ and $\Pcan^*$.
Recall that we assumed that $C_k(\bG)$ can only take integer values, e.g. because it counts certain properties of $\bG$. 
Importantly, we also assume that the parameter value $\vt^*$ conjugate to $\vC^*$ is not a critical point of any phase transition. The possibility that $\vt^*$ approaches a phase boundary is discussed separately elsewhere in the paper.

Using the integral representation $\delta_{x,y}=\int_{-\pi}^{+\pi} \frac{d\psi}{2\pi}\,
e^{i\psi(x-y)}$ of the Kronecker symbol (where $\psi\in[-\pi,+\pi]\subset\R$), and extending it to the $K$-dimensional case, we can write 
\begin{eqnarray}
\Omega_{\vC^*} &=& \sum_{\bG\in \cG}\delta_{\vC^*,\vC(\bG)}\nonumber\\
&=& \sum_{\bG\in \cG} \prod_{k=1}^K\int_{-\pi}^{+\pi} \frac{d\psi_k}{2\pi}\,
e^{i\psi_k[C_k^*-C_k(\bG)]}\nonumber\\
&=& \sum_{\bG\in \cG} \prod_{k=1}^K\int_{\alpha_k}\frac{dz_k}{2\pi i}\,
e^{z_k[C_k^*-C_k(\bG)]},\label{eq:formulazzo1}
\end{eqnarray}
where in the last equality we have rewritten the integral in terms of the complex variable $z_k=\theta_k+i \psi_{k}$ (with $\theta_k$ and $\psi_k$ real) and reinterpreted the real domain of integration $[-\pi,+\pi]\subset\R$ for $\psi_k$ as a segment $\alpha_k\subset\C$ for $z_k$, going from the point of (real, imaginary) coordinates $(0,-\pi)$ to the point $(0,+\pi)$ along the imaginary axis.
This trick, although apparently pointless, allows us to perform a convenient operation, as we now explain.

Note that the function $f_k(z_k)=e^{z_k[C_k^*-C_k(\bG)]}$ being integrated is analytical over the entire complex plane (hence it is an entire function) and its integral along any closed contour is therefore zero:
\be
\oint dz_k f_k(z_k)=0.
\label{eq:contour1}
\ee
For convenience, let us consider the closed rectangular contour formed by the following four segments in the complex plane: the aforementioned upward vertical segment $\alpha_k$ going from $(0,-\pi)$ to $(0,+\pi)$, the horizontal segment $\beta_k$ going from $(0,+\pi)$ to $(\theta^*_k,+\pi)$, the downward vertical segment $\gamma_k$ going from $(\theta^*_k,+\pi)$ to $(\theta^*_k,-\pi)$, and the horizontal segment $\delta_k$ going from $(\theta^*_k,-\pi)$ back to the starting point $(0,-\pi)$.
We can rewrite eq.~\eqref{eq:contour1} for this particular contour as
\begin{equation}
\left[
\int_{\alpha_k}\!\!\!dz_k+
\int_{\beta_k}\!\!\!dz_k+
\int_{\gamma_k}\!\!\!dz_k+
\int_{\delta_k}\!\!\!dz_k
\right]f_k(z_k)=0.
\label{eq:contour2}
\end{equation}
Now, since $C_k(\bG)$ (and hence $C_k^*$) is assumed to take only integer values, we have the following periodicity:
\begin{equation}
f_k(z_k+2\pi i)=e^{(z_k+2\pi i)[C_k^*-C_k(\bG)]}=f_k(z_k).
\end{equation}
This implies that, for any value of $\theta_k$, $f_k$ takes the same value in each pair of points of the type $(\theta_k,-\pi)$, $(\theta_k,+\pi)$. 
Since the segments $\beta_k$ and $\delta_k$ go over such pairs of points in opposite direction, we have
\begin{equation}
\left[\int_{\beta_k}\!\!\!dz_k+
\int_{\delta_k}\!\!\!dz_k
\right]f_k(z_k)=0
\end{equation}
which, together with eq.~\eqref{eq:contour2}, implies
\be
\int_{\alpha_k}\!\!\!dz_kf_k(z_k)=
-\!\int_{\gamma_k}\!\!\!dz_kf_k(z_k)=\!\int_{\omega_k}\!\!\!dz_kf_k(z_k),\label{eq:newdomain}
\ee
where $\omega_k$ is the upward vertical segment going from $(\theta_k^*,-\pi)$ to $(\theta_k^*,+\pi)$, i.e. the same as $\gamma_k$ but going in opposite direction.
Using eq.~\eqref{eq:newdomain}, and denoting with $\vec{\omega}$ the collection of all segments $\{\omega_k\}_{k=1}^K$, eq.~\eqref{eq:formulazzo1} becomes 
\begin{eqnarray}
\Omega_{\vC^*} &=& \sum_{\bG\in \cG} \prod_{k=1}^K\int_{\omega_k}\frac{dz_k}{2\pi i}\,
e^{z_k[C_k^*-C_k(\bG)]}\nonumber\\
&=& \sum_{\bG\in \cG} \int_{\vec{\omega}}\frac{d\vec{z}}{(2\pi i)^K}\,
e^{\vec{z}\cdot[\vC^*-\vC(\bG)]}\nonumber\\
&=& \int_{\vec{\omega}}\frac{d\vec{z}}{(2\pi i)^K}\,
e^{\vec{z}\cdot\vC(\bG^*)}\sum_{\bG\in \cG} e^{-\vec{z}\cdot\vC(\bG)}\nonumber\\
&=& \int_{\vec{\omega}}\frac{d\vec{z}}{(2\pi i)^K}\,e^{\cH(\bG^*,\vec{z})}{\cal Z}(\vec{z})\nonumber\\
&=& \int_{\vec{\omega}}\frac{d\vec{z}}{(2\pi i)^K}\,\Pcan^{-1}(\bG^* | \vec{z}),\nonumber\\
&=& \int_{-\vec{\pi}}^{+\vec{\pi}}\frac{d\vec{\psi}}{(2\pi)^K}\,\Pcan^{-1}(\bG^* | \vt^*+i\vec{\psi}),
\label{eq:formulazzo2}
\end{eqnarray}
where we have extended the domain of $\cH(\bG^*,\vt)$, $\cZ(\vt)$ and $\Pcan(\bG^*|\vt)$, all viewed as functions of $\vt$, from $\R^K$ to $\C^K$. In so doing, they have become complex functions of complex vectors.
To produce the last equality, we have rewritten the integral for $z_k$ along the complex segment $\omega_k\subset\C$ back as an integral for $\psi_k$ along the real interval $[-\pi,+\pi]\subset\R$.
Correspondingly, we have denoted $\pm\vec{\pi}=\pm\pi\vec{1}$, where $\vec{1}$ is the $K$-dimensional vector with all unit entries.

Note that, to be able to interchange the order of sum and integral in the derivation of eq.~\eqref{eq:formulazzo2} even in the limit $n\to\infty$ (which we will take later on), we have to assume uniform convergence of the partial sums of $\sum_{\bG\in \cG} e^{-\vec{z}\cdot\vC(\bG)}$ (which becomes an infinite series in such limit) to $\cZ(\vec{z})$.
Note that our assumptions also ensure that $\cZ(\vec{z})$ is a holomorphic (or analytic) function in a domain containing $\vec{\omega}$. 
For finite $n$, this is automatically ensured by the fact that $\cZ(\vec{z})$ is a finite sum of exponentials (which are all holomorphic). In the thermodynamic limit $n\to\infty$, the analyticity of $\cZ(\vec{z})$ is ensured by our assumption that  $\vec{\omega}$ does not cross any boundary between different thermodynamic phases, i.e. that the point $\vt^*$, which is the only real point crossed by $\vec{\omega}$, is not a critical point of any phase transition.

Recalling from eq.~\eqref{eq:Pmic} that $\Pmic(\bG^*|\vC^*)=\Omega^{-1}_{\vC^*}$, eq.~\eqref{eq:formulazzo2} immediately proves eq.~\eqref{eq:connection} in the main text, which provides an exact relationship between the microcanonical and the canonical probability extended to the complex domain. Remarkably, this relationship is valid even for finite $n$ and even if EE does not hold.

\subsection*{Asymptotic relation between $\Omega^*$ and the canonical (co)variances}
Here we prove another main result, i.e. we start from eq.~\eqref{eq:formulazzo2} and, using a saddle point calculation, arrive at its asymptotic expression given in eq.~\eqref{eq:omegaapprox}.
We keep assuming that $\vt^*$ is not a critical point of any phase transition.

First, extending the log-likelihood function to the complex domain via the definition
\be
\cL^*(\vec{z})\equiv\ln \Pcan(\bG^* |\vec{z}),
\ee
we rewrite eq.~\eqref{eq:formulazzo2} as
\be
\Omega_{\vC^*}
= \int_{\vec{\omega}}\frac{d\vec{z}}{(2\pi i)^K}\,e^{-\cL^*(\vec{z})}. \label{eq:formulazzo3}
\ee

Second, we note that $\cL^*(\vec{z})$ is a holomorphic function, which follows from the analyticity of $\Pcan(\bG^*|\vec{z})$ away from phase transitions and from the fact that, when $\Pcan(\bG^*|\vec{z})$ is real-valued, it is positive and hence its logarithm is analytic.

Third, we notice that the point $\vec{z}^*\in\vec{\omega}$ of coordinates $(\vt^*,\vec{0})$ is a saddle point for $\cL^*(\vec{z})$.
To see this, we extend the calculation in eq.~\eqref{eq:gradient} to the complex domain to get 
\be
\frac{\partial}{\partial z_k}\cL^*(\vec{z})=\langle C_k|\vec{z}\rangle-C^*_k
\label{eq:zgradient}
\ee
and note that eq.~\eqref{eq:nogradient} implies
\be
\frac{\partial}{\partial z_k}\cL^*(\vec{z})\bigg|_{\vec{z}=\vec{z}^*}=\langle C_k|\vt^*\rangle-C^*_k=0\quad\forall k.
\label{eq:nozgradient}
\ee
The above result shows that $\vec{z}^*$ is a stationary point for $\cL^*(\vec{z})$. As always happens in the complex domain, such a point is necessarily a saddle point. This can be confirmed here by noting that, as we move across $\vec{z}^*$ along the real direction, the second derivatives are given by eq.~\eqref{eq:sigma2} and the corresponding Hessian matrix is, as already mentioned, negative semi-definite.
On the contrary, as we show later, as we move across $\vec{z}^*$ along the imaginary direction, all second derivatives have the opposite sign and the Hessian matrix is therefore positive semi-definite.
The sign difference of the second derivatives calculated along the real and imaginary directions implies that $\vec{z}^*$ is a saddle point.

Fourth, we notice that $\vec{z}^*$ is the only relevant saddle point for evaluating eq.\eqref{eq:formulazzo3}.
Due to the periodicity $\cL^*(\vec{z})=\cL^*(\vec{z}+2\pi i\vec{u})$ where $\vec{u}$ is any vector in $\Z^K$, all the points of the type $\vec{z}^*+2\pi i\vec{u}$, having coordinates $(\vt^*,2\pi \vec{u})$, are also saddle points for $\cL^*(\vec{z})$. 
However, among all these saddle points, $\vec{z}^*$ is the only one encountered along the path $\vec{\omega}$. 
It is then reasonable to assume that there are no other saddle points along $\vec{\omega}$, i.e. no other points $(\vt^*,\vec{\psi})$ with $\vec{\psi}\ne 0$ such that $\langle \vec{C}|\vt^*+i\vec{\psi}\rangle=\vC^*$ (the existence of such points is unlikely, given that $\vC^*\in\R^K$).
At the very least, we assume that those points, if present, can be avoided via a suitable deformation of $\vec{\omega}$ (since $\cL^*(\vec{z})$ is analytical, such deformation is possible, unless those points form an infinite curve separating $\vec{z}^*$ from the endpoints $\pm\vec{\pi}$ of integration, or a closed curve surrounding $\vec{z}^*$ and/or $\pm\vec{\pi}$).

The above considerations imply that, in the thermodynamic limit, eq.~\eqref{eq:formulazzo3} naturally lends itself to a saddle-point calculation~\cite{saddlepoint} leading to eq.~\eqref{eq:omegaapprox}. We now show this in detail. We define the complex-valued function
\be
\cU^*(\vec{\psi})=\cL^*(\vt^*+i\vec{\psi}),\quad\cU^*:\R^K\to\C
\label{eq:U}
\ee
and Taylor-expand it around $\vec{\psi}=\vec{0}$:
\begin{eqnarray}
\cU^*(\vec{\psi})\!\!&=&\!\!\!\!\sum_{m=0}^{\infty}\frac{1}{m!}  \sum_{\sum_k\!a_k=m}\!
\frac{\partial^m}{\partial \psi_1^{a_1}
\dots \partial\psi_K^{a_K}} \cU(\vec{\psi}\,)\bigg|_{\vec{\psi}=\vec{0}}\prod_{k=1}^K\psi_k^{a_k}\nonumber\\
&=&\cU^*(\vec{\psi})\bigg|_{\vec{\psi}=\vec{0}}+\sum_{k=1}^K\psi_k\frac{\partial}{\partial\psi_k}\cU^*(\vec{\psi})\bigg|_{\vec{\psi}=\vec{0}}\nonumber\\
&&+\frac{1}{2}\sum_{k,l}\psi_k\psi_l\frac{\partial^2}{\partial\psi_k\partial\psi_l}\cU^*(\vec{\psi})\bigg|_{\vec{\psi}=\vec{0}}+{\cal V}^*(\vec{\psi}),
\label{eq:Taylor1}
\end{eqnarray}
where ${\cal V}^*(\vec{\psi})=O(|\vec{\psi}|^3)$ contains all the higher-order terms and is such that ${\cal V}^*(\vec{0})=0$.
Note that the function $\cU^*$ has the periodicity $\cU^*(\vec{\psi})=\cU^*(\vec{\psi}+2\pi \vec{u})$ for any $\vec{u}\in\Z^K$, therefore the above expansion only makes sense within one such period.  
This is consistent with the fact that, in eq.~\eqref{eq:formulazzo2}, the domain of integration for each variable $\psi_k$ is the single period $[-\pi,+\pi]$.
Now, from the identity
\begin{equation}
\label{eq:connect}
\frac{\partial^m}{\partial \psi_1^{a_1}\dots \partial\psi_K^{a_K}}\, \cU^*(\vec{\psi}\,)\bigg|_{\vec{\psi}=\vec{0}}
\!\!= i^m\,\frac{\partial^m}{\partial \theta_1^{a_1}\dots\partial\theta_k^{a_k}}\,\cL^*(\vec{\theta}\,)
\bigg|_{\vec{\theta}=\vec{\theta^*}}\nonumber
\end{equation}
we realize that all the $m$-th order derivatives of $\cU^*(\vec{\psi})$ at $\vec{\psi}=\vec{0}$ are real if $m$ is even and purely imaginary if $m$ is odd, because all the derivatives of $\cL^*(\vec{\theta})$, when evaluated at $\vt=\vt^*$, are real.
Moreover, we can use eqs.~\eqref{eq:nogradient},~\eqref{eq:sigma2} and \eqref{eq:L=-S} to obtain
\begin{eqnarray}
\cU^*(\vec{\psi})\bigg|_{\vec{\psi}=\vec{0}}\!\!&=&\!\!\cL^*(\vt)\bigg|_{\vec{\theta}=\vec{\theta^*}}=-\Scan^*,\\
\frac{\partial}{\partial\psi_k}\cU^*(\vec{\psi})\bigg|_{\vec{\psi}=\vec{0}}\!\!&=&\!\!i\frac{\partial}{\partial\theta_k}\cL^*(\vt)\bigg|_{\vt=\vt^*}\!\!=0,\label{eq:U'}\\
\frac{\partial^2}{\partial\psi_k\partial\psi_l}\cU^*(\vec{\psi})\bigg|_{\vec{\psi}=\vec{0}}\!\!&=&\!\!-\frac{\partial^2}{\partial\theta_k\partial\theta_l}\cL^*(\vt)\bigg|_{\vt=\vt^*}\!\!=\Sigma^*_{k,l}.\label{eq:U''}\quad
\end{eqnarray}
Equation~\eqref{eq:U''} confirms that $\vec{z}^*$ is a stationary point for $\cL^*(\vec{z})$, as anticipated.

The expansion in eq.~\eqref{eq:Taylor1} can then be rewritten as
\begin{eqnarray}
\cU^*(\vec{\psi})\!&=&-\Scan^*+\frac{1}{2}\sum_{k,l}\psi_k\psi_l\Sigma^*_{k,l}+{\cal V}^*(\vec{\psi})
\label{eq:Taylor2}
\end{eqnarray}
and, via eq.~\eqref{eq:U}, we can rewrite eq.~\eqref{eq:formulazzo3} as
\begin{eqnarray}
\Omega_{\vC^*}
&=& \int_{-\vec{\pi}}^{+\vec{\pi}}\!\!\frac{d\vec{\psi}}{(2\pi)^K}\,e^{-\cL^*(\vt^*+i\vec{\psi})} \nonumber\\
&=& \int_{-\vec{\pi}}^{+\vec{\pi}}\!\!\frac{d\vec{\psi}}{(2\pi)^K}\,e^{-\cU^*(\vec{\psi})},\nonumber\\
&=& e^{\Scan^*}\!\!\int_{-\vec{\pi}}^{+\vec{\pi}}\!\!\frac{d\vec{\psi}}{(2\pi)^K}\,e^{-\frac{1}{2}\vec{\psi}\,\bS^*\vec{\psi}} \,e^{-{\cal V}^*(\vec{\psi})}.
\label{eq:formulazzo4}
\end{eqnarray}
We are now ready for a saddle-point calculation~\cite{saddlepoint}. Rather than using the multidimensional version of the method, which requires the existence of a common large factor at the exponent for all the coordinates of $\vec{\psi}$, we prefer iterating the one-dimensional version upon diagonalizing $\bS^*$, so that we can keep track of the asymptotic behaviour of each of its eigenvalues separately.
Let us therefore carry out a volume-preserving (i.e. with unit Jacobian determinant) linear change of variables $\vec{\psi}\mapsto\vec{\xi}(\vec{\psi})$ that diagonalizes $\bS^*$ while keeping $\vec{\xi}(\vec{0})=\vec{0}$ and rewrite
\be
\Omega_{\vC^*}
= e^{\Scan^*}\!\!\int_{\vec{\xi}^-}^{\vec{\xi}^+}\!\!\frac{d\vec{\xi}}{(2\pi)^K}\frac{e^{-\frac{1}{2}\sum_{k=1}^K\lambda^*_k\xi_k^2}}{e^{{\cal W}^*(\vec{\xi})}},
\label{eq:formulazzo5}
\ee
where $\lambda^*_k$ is the $k$-th eigenvalue of $\bS^*$, ${\cal W}^*(\vec{\xi})={\cal V}^*[\vec{\psi}(\vec{\xi})]$ (so that ${\cal W}^*(\vec{0})={\cal V}^*(\vec{0})=0$), and $\vec{\xi}^{\pm}=\vec{\xi}(\pm\vec{\pi})$.

As discussed previously, the $K$ eigenvalues of $\bS^*$ cannot be negative.
We already mentioned that if only $\tilde{K}\le K$ constraints are mutually independent, then $K-\tilde{K}$ eigenvalues will be zero. 
Let us now carry out the integration in eq.~\eqref{eq:formulazzo5} over a single variable $\xi_l$ for which the corresponding eigenvalue $\lambda^*_l$ is zero.
In such a case, eq.~\eqref{eq:Taylor2} implies that, along the direction $\xi_l$, the function $\cU^*[\vec{\psi}(\vec{\xi})]$ is constant and equal to the value $\cU^*[\vec{\psi}(\vec{\xi})]\big|_{\xi_l=0}$.
This means that, along $\xi_l$, ${\cal W}^*(\vec{\xi})$ is also constant and equal to ${\cal W}^*(\vec{\xi})\big|_{\xi_l=0}$.
Therefore, if $\lambda^*_l=0$ then in eq.~\eqref{eq:formulazzo5} the integral over $\xi_l$ is the definite integral of a constant function, which generates a contribution $\xi_l^+-\xi_l^-=2\pi$. This follows from the fact that the trasformation $\vec{\psi}\mapsto\vec{\xi}$ is volume-preserving, so it should keep the area along each redundant direction equal to the original area $2\pi$ (because each such direction can be added arbitrarily without changing the volume spanned by the non-redundant directions).
This contribution effectively replaces the factor $(2\pi)^K$ at the denominator with $(2\pi)^{K-1}$.
\be
\Omega_{\vC^*}
= e^{\Scan^*}\!\!\int_{\vec{\xi}^-_{-l}}^{\vec{\xi}^+_{-l}}\!\!\frac{d\vec{\xi}_{-l}}{(2\pi)^{K-1}}\frac{e^{-\frac{1}{2}\sum_{k\ne l}\lambda^*_k\xi_k^2}}{e^{{\cal W}^*(\vec{\xi})}\big|_{\xi_l=0}},
\label{eq:formulazzozero}
\ee
where $\vec{\xi}_{-l}$ denotes the $(K-1)$-dimensional vector including all variables except $\xi_l$.
Repeating the procedure for all the zero eigevalues, the factor $(2\pi)^{K-1}$ becomes $(2\pi)^{\tilde{K}}$.

Let us now start again from eq.~\eqref{eq:formulazzo5} and consider an eigenvalue $\lambda^*_l$ that is strictly positive.
The corresponding term $e^{-\frac{1}{2}\lambda^*_l\xi_l^2}$ in the integral is now an unnormalized Gaussian function with zero mean and variance $1/\lambda^*_l$. 
If $\lambda^*_l$ grows with $n$, the Gaussian function $e^{-\frac{1}{2}\lambda^*_l\xi_l^2}$ will become more and more peaked aroud zero as $n\to\infty$ and its limit becomes the Dirac delta function $\delta(\xi_l)$ times the normalization factor $\sqrt{2\pi/\lambda^*_l}$.
Integrating over the variable $\xi_l$, for $n\to\infty$ the limiting $\delta(\xi_l)$ will select the unique value $e^{-{\cal W}^*(\vec{\xi})}\big|_{\xi_l=0}$ in eq.~\eqref{eq:formulazzo5}.
If $\lambda^*_l$ remains finite (and positive) as $n$ grows, there is an additional correction factor that can be estimated~\cite{saddlepoint} as $\left[1+O\left(1/\lambda^*_l\right)\right]$.
So in general we get
\be
\Omega_{\vC^*}\!
= \frac{e^{\Scan^*}}{\sqrt{\lambda^*_l}}\!\int_{\vec{\xi}^-_{-l}}^{\vec{\xi}^+_{-l}}\!\!\!\!\frac{d\vec{\xi}_{-l}}{(2\pi)^{K-1/2}}\frac{e^{-\frac{1}{2}\sum_{k\ne l}\lambda^*_k\xi_k^2}}{e^{{\cal W}^*(\vec{\xi})}\big|_{\xi_l=0}}\left[1+O\left(1/\lambda^*_l\right)\right]
\label{eq:formulazzofinite}
\ee
where $\vec{\xi}_{-l}$ denotes again the $(K-1)$-dimensional vector including all variables except $\xi_l$.

Iterating the above calculations for all eigenvalues, and using either eq.\eqref{eq:formulazzozero} (whenever $\lambda^*_l=0$) or eq.~\eqref{eq:formulazzofinite} (whenever $\lambda^*_l>0$), we arrive at the asymptotic expansion
\begin{eqnarray}
\Omega_{\vC^*}&
=& \frac{e^{\Scan^*}}{\sqrt{\lambda^*_1\cdots\lambda^*_{\tilde{K}}}}
\frac{\prod_{k=1}^{\tilde{K}}\left[1+O\left(1/\lambda^*_k\right)\right]}{(2\pi)^{\tilde{K}-\tilde{K}/2}\,\,e^{{\cal W}^*(\vec{0})}}\nonumber\\
&=&
\frac{e^{\Scan^*}}{\sqrt{(2\pi)^{\tilde{K}}\det\tilde{\bS}^*}}\prod_{k=1}^{\tilde{K}}\left[1+O\left(1/\lambda^*_k\right)\right]\nonumber\\
&=&
\frac{e^{\Scan^*}\,T^*}{\sqrt{\det(2\pi\tilde{\bS}^*)}},
\label{eq:proof}
\end{eqnarray}
where $\tilde{\bS}^*$ is the reduced ${\tilde{K}\times\tilde{K}}$ matrix of the canonical (co)variances of the ${\tilde{K}\le K}$ independent constraints,
and we have reordered the original $K$ constraints in such a way that the first $\tilde{K}$ ones are independent (so the first $\tilde{K}$ eigenvalues are non-zero).
The above derivation proves eq.~\eqref{eq:omegaapprox}.
The asymptotic behaviour of $T^*$ as a function of that of the eigenvalues of $\tilde{\bS}^*$ is discussed later.

\subsection*{Exact calculation of $\Delta^*$}
We now use the previous results in order to calculate the relative entropy $\Delta^*$ explicitly.
Inserting eq.~\eqref{eq:formulazzo2} into eq.~\eqref{eq:Qcan} implies
\begin{eqnarray}
\Delta^*&=&-\ln\int_{-\vec{\pi}}^{+\vec{\pi}}\!\! \frac{d\vec{\psi}}{(2\pi)^K}\,\frac{\Pcan(\bG^*|\vt^*)}{\Pcan(\bG^* | \vt^*+i\vec{\psi})}\nonumber\\
&=&-\ln\int_{-\vec{\pi}}^{+\vec{\pi}}\!\! \frac{d\vec{\psi}}{(2\pi)^K}\,\frac{e^{-\cH(\bG^*,\vt^*)}}{\cZ(\vt^*)}\frac{\cZ(\vt^*+i\vec{\psi})}{e^{-\cH(\bG^*,\vt^*+i\vec{\psi})}}\nonumber\\
&=&-\ln\int_{-\vec{\pi}}^{+\vec{\pi}}\!\! \frac{d\vec{\psi}}{(2\pi)^K}\,e^{-\vt^*\cdot\vC^*+\vt^*\cdot\vC^*+i\vec{\psi}\cdot\vC^*}\frac{\cZ(\vt^*+i\vec{\psi})}{\cZ(\vt^*)}\nonumber\\
&=&-\ln\int_{-\vec{\pi}}^{+\vec{\pi}}\!\! \frac{d\vec{\psi}}{(2\pi)^K}\,e^{i\vec{\psi}\cdot\vC^*}\frac{\cZ(\vt^*+i\vec{\psi})}{\cZ(\vt^*)}\nonumber\\
&=&-\ln\int_{-\vec{\pi}}^{+\vec{\pi}}\!\! \frac{d\vec{\psi}}{(2\pi)^K}\,e^{i\vec{\psi}\cdot\vC^*}\langle e^{-i\vec{\psi}\cdot\vC}|\vt^*\rangle\nonumber\\
&=&-\ln\int_{-\vec{\pi}}^{+\vec{\pi}}\!\! \frac{d\vec{\psi}}{(2\pi)^K}\,\langle e^{i\vec{\psi}\cdot(\vC^*-\vC)}|\vt^*\rangle
\label{eq:provephi}
\end{eqnarray}
where we have used the definition~\eqref{eq:ensembleaverage} of ensemble average of the (complex) function $X(\bG)=e^{i\vec{\psi}\cdot[\vC^*-\vC(\bG)]}$ and exploited the fact that
\begin{eqnarray}
\frac{\cZ(\vt^*+i\vec{\psi})}{\cZ(\vt^*)}&=&\frac{\sum_\bG e^{-(\vt^*+i\vec{\psi})\cdot\vC(\bG)}}{\sum_\bG e^{-\vt^*\cdot\vC(\bG)}}\nonumber\\
&=&\sum_\bG \Pcan(\bG|\vt^*)e^{-i\vec{\psi}\cdot\vC}\nonumber\\
&=&\langle e^{-i\vec{\psi}\cdot\vC}|\vt^*\rangle
\end{eqnarray}
Equation~\eqref{eq:provephi} proves eq.~\eqref{eq:Phigeneral} in the main text.
If the integral can be carried out explicitly, eq.~\eqref{eq:provephi} allows an exact calculation of $\Delta^*$ in terms of only the canonical probability $\Pcan^*$, thus avoiding any microcanonical enumeration.

\section*{Asymptotic behaviour of $T^*$}
We now study the asymptotic behaviour of the correction term $T^*$ appearing in eqs.~\eqref{eq:omegaapprox} and~\eqref{eq:deltaT}.

We already noted that, in general, some of the nonzero eigenvalues of $\tilde{\bS}^*$ may diverge, while some may remain finite, as $n\to\infty$. 
If there are $K_\infty\le\tilde{K}$ diverging eigenvalues, and if we choose an ordering such that these eigenvalues are the first ones, then their contribution to $T^*$ is
\be
\prod_{k=1}^{K_{\infty}}\left[1+O\left(1/\lambda^*_k\right)\right]=1+O\bigg(\sum_{k=1}^{K_{\infty}}1/\lambda^*_k\bigg)=1+O(K_{\infty}/\lambda^*_{\infty})
\label{eq:Tinfinite}
\ee
where $\lambda^*_{\infty}$ is the (diverging) harmonic mean of the $K_\infty$ diverging eigenvalues:
\be
\frac{1}{\lambda^*_{\infty}}=\frac{1}{K_{\infty}}\sum_{k=1}^{K_\infty}\frac{1}{\lambda^*_k}.
\ee
The remaining $\tilde{K}-K_\infty$ finite eigenvalues give a contribution to $T^*$ equal to 
\begin{eqnarray}
\prod_{k=K_{\infty}+1}^{\tilde{K}}\left[1+O\left(1/\lambda^*_k\right)\right]&=&\prod_{k=K_\infty+1}^{\tilde{K}}O\big(1/\lambda^*_k\big)\nonumber\\
&=&O\left((\bar{\lambda}^*)^{-(\tilde{K}-K_{\infty})}\right),
\label{eq:Tfinite}
\end{eqnarray}
where
\be
\bar{\lambda}^*\equiv\bigg(\prod_{k=K_\infty+1}^{\tilde{K}}\lambda^*_k\bigg)^{1/(\tilde{K}-K_{\infty})}
\ee
is the (finite) geometric mean of the $\tilde{K}-K_{\infty}$ finite eigenvalues.

Combining eqs.~\eqref{eq:Tinfinite} and~\eqref{eq:Tfinite}, we arrive at 
\be
T^*=\big[1+O(K_{\infty}/\lambda^*_{\infty})\big]O\big[(\bar{\lambda}^*)^{-(\tilde{K}-K_{\infty})}\big]
\ee
and, taking the logarithm, we get
\be
\ln T^*=\ln\big[1+O(K_{\infty}/\lambda^*_{\infty})\big]+O(\tilde{K}-K_{\infty}).
\label{eq:lnT}
\ee
Correspondingly, the quantity $\tau^*_{\alpha_\infty}$ defined in eq.~\eqref{eq:tau} becomes
\be
\tau^*_{\alpha_\infty}=-\lim_{n\to\infty}\frac{\ln\big[1+O(K_{\infty}/\lambda^*_{\infty})\big]+O(\tilde{K}-K_{\infty})}{\alpha_n}
\label{eq:tau2}
\ee
The above equation is very important as it informs us about the limiting behaviour of $\delta^*_{\alpha_\infty}$, as we now discuss. 

Let us first consider the case when all the nonzero eigenvalues are diverging, i.e. $K_{\infty}=\tilde{K}$.
Note that this is the typical situation, since the constraints are generally macroscopic quantities whose values, as well as those of their fluctuations (hence the eigenvalues of the covariance matrix), diverge at least like $n$ in the thermodynamic limit.
If the number of constraints remains finite as $n\to\infty$  (i.e. $K_{\infty}<\infty$), then $\ln T^*=O(1/\lambda^*_{\infty})\to0$ and $\Delta^*\to\ln\sqrt{\det(2\pi\tilde{\bS}^*)}$. Stated more rigorously, for any increasing $\alpha_n$ we get $\tau^*_{\alpha_\infty}=0$.
If all eigenvalues are diverging, and the number $K_{\infty}$ of constraints diverges like $\lambda^*_{\infty}$, $\ln T^*$ has now a finite limit but still vanishes upon dividing by $\alpha_n$. We therefore still obtain $\tau^*_{\alpha_\infty}=0$. The same holds true as long as $K_{\infty}$ diverges slower than the astronomically big number $\lambda^*_{\infty}e^{\alpha_n}$.
Therefore $\tau^*_{\alpha_\infty}=0$ whenever all eigenvalues diverge. 

In the opposite case when all eigenvalues have a finite limit as $n\to\infty$ (i.e. $K_{\infty}=0$), eq.~\eqref{eq:lnT} implies $\ln T^*=O(\tilde{K})$. Upon dividing by $\alpha_n$ and taking the limit, this term vanishes, so $\tau^*_{\alpha_\infty}=0$ still holds, as long as the number $\tilde{K}$ of constraints is bounded or diverges slower than $\alpha_n$. By contrast, if the number of constraints grows like $\alpha_n$ or faster, $\tau^*_{\alpha_\infty}>0$ and its value has to be evaluated case by case. 

Combining all the results discussed so far, we conclude that
eq.~\eqref{eq:deltalimit} with $\tau^*_{\alpha_\infty}=0$ always holds as an exact result, unless $\bS^*$ has an infinite number (growing at least like $\alpha_n$) of finite eigenvalues, in which case $\tau^*_{\alpha_\infty}>0$ and has to be calculated for the specific model under scrutiny.

\clearpage
\end{document}